\documentclass[twocolumn]{elsart3}

\usepackage{graphicx}

\usepackage{amssymb}

\begin{document}
  
\begin{frontmatter}



\title{Validation Studies of the ATLAS Pixel Detector Control System}


\author[WUP]{Joachim Schultes}, \author[WUP]{Karl-Heinz Becks}, 
\author[WUP]{Tobias Flick}, \author[WUP]{Tobias Hen\ss}, 
\author[WUP]{Martin Imh\"{a}user}, \author[WUP]{Susanne Kersten},  
\author[WUP]{Peter Kind}, \author[WUP]{Kerstin Lantzsch}, 
\author[WUP]{Peter M\"{a}ttig}, \author[WUP]{Kendall Reeves}, and 
\author[BONN]{Jens Weingarten}

\address[WUP]{University of Wuppertal, Gau\ss str. 20, 42097 Wuppertal, Germany}
\address[BONN]{University of Bonn, Nussallee 12, 53115 Bonn, Germany}

\begin{abstract}
  The ATLAS pixel detector consists of 1744 identical silicon pixel modules 
arranged in three barrel  layers providing coverage for the central region,
and three disk layers on either side of the primary interaction point providing
coverage of the forward regions. 

Once deployed into the experiment, the detector will employ optical data 
transfer, with the requisite powering being provided by a complex system 
of commercial and custom-made power supplies.  However, during normal 
performance and production tests in the laboratory, only single modules are 
operated and electrical readout is used.  In addition, standard laboratory 
power supplies are used.

In contrast to these normal tests, the data discussed here was obtained from a 
multi-module assembly which was powered and read out using production items: 
the optical data path, the final design power supply system using close to 
final services, and the Detector Control System (DCS).

To demonstrate the functionality of the pixel detector system a stepwise 
transition was made from the normal laboratory readout and power supply 
systems to the ones foreseen for the experiment, with validation of the data 
obtained at each transition.
\end{abstract}

\begin{keyword}
ATLAS \sep Pixel \sep Detector Control System (DCS) \sep System Test 
\sep Power Supplies \sep Interlock System
\end{keyword}
\end{frontmatter}

\section{The Atlas Pixel Detector}
\label{PixDet}

The pixel detector of the ATLAS experiment is the innermost part of the
inner detector, and will provide crucial information for precise vertex
determination.  It will consist of 6 disks, 3 on each side of the 
interaction region, and 3 barrel layers. The disks are comprised of 8 
sectors, each of which is equipped with 6 detector modules.  The 3 barrel
layers are composed of bi-staves (each one equipped with 26 detector modules). 
The layer closest to the beam pipe (referred to as the B-Layer) will have 11
bi-staves, while the intermediate layer (Layer 1) and the outer layer 
(Layer 2) will have 19 and 26 of these bi-staves, respectivley. 
A bi-stave is divided into 4 half-staves, each with 6 or 7 detector modules.

In total the pixel detector will have 1744 detector modules possessing 
46080 pixel cells each, resulting in a total number of just over 80 million 
readout channels.

\begin{figure}[htb]
  \includegraphics[width=\columnwidth]{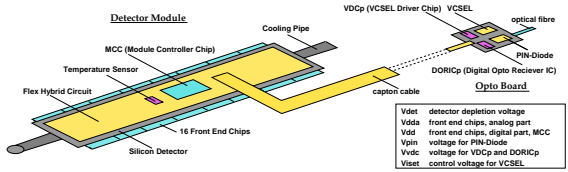}
  \caption{Scheme of detector module}
  \label{fig:Module}
\end{figure}

Each detector module (Figure \ref{fig:Module}) consists of 16 frontend 
chips which are connected to the associated sensor cells. The readout and 
control of the frontend chips is provided by the Module Controller Chip (MCC). 
The analog part of the frontend chips is separately supplied, while their 
digital part is supplied together with the MCC.

During operation of a detector module, it is also necessary to provide 
a depletion supply channel for the sensor. In addition, the detector module 
will have a thermistor (NTC\footnote{NTC: \textbf{n}egative 
  \textbf{t}emperature \textbf{c}oefficient}) which is used to provide 
temperature information for each detector module. This thermistor,  
with a related interlock circuit \cite{Interlock}, is used to prevent damage 
to the hardware due to high temperatures. 

\section{The Read out Chain}
\label{Readout}
For the data stream between the detector and the read out system an 
optical transfer for most of the distance of about 80\,m  will be used.

An opto board is used on the on-detector side to convert the electrical
signal into optical signals and vice versa (Figure \ref{fig:PSS}). The supply 
of this opto board is handled by the SC-OLink\footnote{SC-OLink: 
  \textbf{S}upply and \textbf{C}ontrol for the \textbf{O}pto\textbf{Link}} 
which is, like all of the supply components, placed outside the 
detector volume.

The control of the detector modules, the data taking, and the 
histogramming of the data is handled by a ROD
\footnote{ROD: \textbf{R}ead\textbf{o}ut \textbf{D}river}. On its back, 
a BOC\footnote{BOC: \textbf{B}ack \textbf{o}f \textbf{C}rate card} picks up 
the optical transmitter Tx, the receiver Rx, and the S-Link card. Inside the 
counting room, up to 16 ROD/BOC combinations are placed inside a VME crate 
together with a single board computer which handles the data stream.

\section{The Embedded Local Monitor Board}
\label{ELMB}
The ELMB\footnote{\textbf{E}mbedded \textbf{L}ocal \textbf{M}onitor 
  \textbf{B}oard}, developed by the ATLAS DCS group, is a multi purpose, low 
cost I/O device to monitor and control various hardware components 
\cite{ELMB}. Each ELMB provides up to 64 multiplexed ADC channels 
and 24 digital I/O-lines. Its CAN\footnote{\textbf{C}ontroller \textbf{A}rea 
  \textbf{N}etwork} bus interface and an OPC\footnote{\textbf{O}LE for 
  \textbf{P}rocess \textbf{C}ontrol} server provided by the ATLS DCS group 
allow the integration into the higher level DCS software. 

One group of required channels in the pixel DCS accounts for the 
large number of temperature sensors that are used throughout the detector 
volume. The majority of those is formed by the detector module thermistors. 
Additionally the monitoring of the environment and of other temperature 
sensitive components needs more ADC channels. All devices are equipped with 
10\,k$\Omega$ negative temperature coefficient thermistors and their ead out 
is based on the use of the ELMB. 
The control of our custom made DCS hardware is based on the usage of the 
ELMB (see below).

\section{Power Supply System}
\label{PSS}
\begin{figure}[htb]
  \includegraphics[angle=-90, width=\columnwidth]{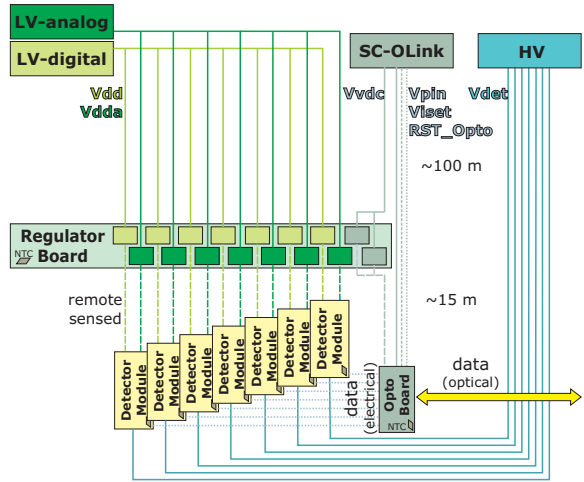}
  \caption{Powering scheme of a half stave or a disc sector}
  \label{fig:PSS}
\end{figure}

The power supply system has to provide floating channels to be 
compatible to the ATLAS grounding scheme. Additionaly the outputs should be 
adjustable over the full range. A high granularity is aimed for to keep the 
number of operating elements as high as possible.

To provide the supply of the detector modules, commercial power 
supplies will be used. They will be placed outside the detector volume 
(off-detector) in the ATLAS counting rooms. Therefore, the services need to 
be longer than 100\,m, and additional devices are required to provide an 
regulated input for the low voltage channels with higher current to the 
devices (Figure \ref{fig:PSS}).

To deplete the sensors an iseg\footnote{Rossendorf, Germany} 
EHQ F007n-F with 16 outputs will be used. In the beginning of the experiment 
one channel will supply 6 or 7 detector modules of a sector or a half-stave. 
In case of increased leakage current inside the sensors due to radiation 
damages, the modularity will be reduced during the lifetime of the experiment.

For the low voltages (analog and digital) a WIENER\footnote{Burscheid, 
  Germany} power supply will provide these, using 2 channels to supply 6 or 7 
detector modules (corresponding again to a sector or a half-stave). An active 
regulation station \cite{IEEE2004_PSS} inside the detector volume will be used 
to regulate the 2\,x\,6 or 2\,x\,7 needed voltages for the whole 
sector or half-stave using remote sensing.\\

\subsection{The Supply and Control for the OptoLink}
\label{PSS_SC-OLink}
\begin{figure}[htb]
  \includegraphics[width=\columnwidth]{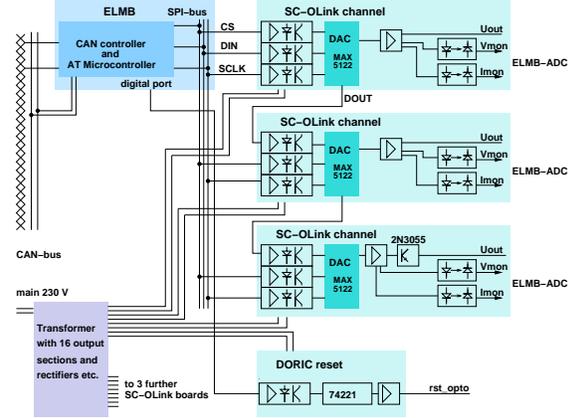}
  \caption{Schematic of the SC-OLink}
  \label{fig:SC-OLink}
\end{figure}
Supplying the opto board will be done by the SC-OLink developed by the 
Wuppertal DCS group \cite{IEEE2004_PSS}, each opto board having its separate 
power supply outputs. Its design is based on the use of the ELMB whichs allows 
to control a high number of channels economically priced.

The SC-OLink provides two low current channels (20\,V and 5\,V, 20\,mA 
each), one 10\,V 800\,mA channel, and one reset signal. Each of the output 
channels uses a separate transformer input to achieve a galvanic separation. 
The monitoring of the voltages and currents are separated from the measurement 
circuit by using linear opto couplers. The precision for the channels is 
better than 8\,bit for the output as well as for the monitored values. On the 
digital side the used DACs\footnote{DAC: \textbf{D}igital \textbf{A}nalog 
  \textbf{C}onverter} MAX 5122 are controlled by a 
SPI\footnote{SPI: \textbf{S}erial \textbf{P}eripheral \textbf{I}nterface, 
  synchronous serial bus}, each one is separated through opto couplers.\\
The 10\,V 800\,mA channel does not use remote sensing, as it will be 
adjusted by the regulator station using two regulators for redundancy.

\subsection{The Regulator Station}
\label{PSS_Regulator}
To protect the frontend chips, which are fabricated in deep-submicron 
technology, against transients, remotely programmable regulator stations 
are installed as close to the detector modules as possible. This radiation 
hard system has been developed by the INFN Milano group\cite{IEEE2004_PSS}. 
The regulators compensate for the large voltage drops on the low mass cables 
in the detector active volume. In parallel they provide an individually 
adjustable control of the low voltage lines for each detector module. The 
core of the system is an ST regulator LHC4913 from ST Microelectronics 
\footnote{Catania, Italy}, which can provide a maximum current up to 3\,A 
and accepts input voltages up to 14\,V.  Using digital trimmers, the 
output voltages can be adjusted. The control is based on an FPGA 
(XC4036XLA-09HQ240C) from Xilinx\footnote{San Jos\'e, USA}, while the 
communication to the outer world is established by ELMBs.

\section{The Detector Control System}
The DCS\footnote{DCS: \textbf{D}etector \textbf{C}ontrol \textbf{S}ystem} is 
based on the development environment PVSS II of the Austrian company 
ETM\footnote{Eisenstadt, Austria} and will be used in all LHC experiments 
to build up the SCADA\footnote{\textbf{S}upervisory \textbf{C}ontrol 
\textbf{A}nd \textbf{D}ata \textbf{A}cquisition} systems. It enables 
the developer to establish all necessary connections to the supply and 
protection system. To give the shifters an overview over the system status, 
and to allow them to operate the system, the DCS provides an easy to use 
graphical user interface.
	
\label{DCS}
\begin{figure}[htb]
  \includegraphics[width=\columnwidth]{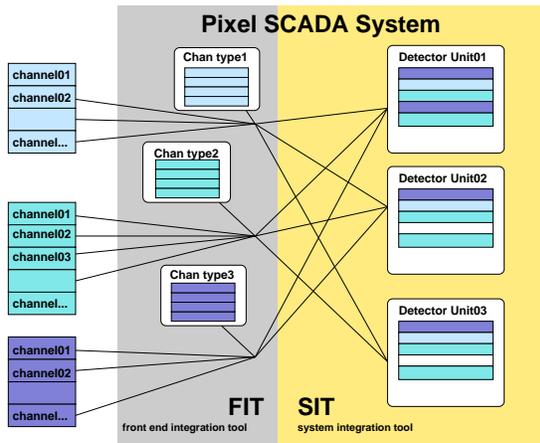}
  \caption{Scheme of the mapping between the channels provided by the 
Frontend Integration Tool to the detector units managed by the System 
Integration Tool}
  \label{fig:FIT-SIT}
\end{figure}
The core of the detector control system consists of two major parts 
(Figure \ref{fig:FIT-SIT}): front end integration (power supplies, 
sensors...) and detector integration (detctor modules, half staves...). 
It is supplemented by the communication with the data taking system and 
a finite state machine to simplify the control, which is currently under 
development. 

\subsection{The Frontend Integration Tools}
\label{FIT}	
To establish the connection to the front end hardware, the FITs
\footnote{FIT: \textbf{F}rontend \textbf{I}ntegration \textbf{T}ool} 
are used. Besides being the underlying layer for all other hardware 
DCS components, the FIT also provides panels to monitor and control 
the connected front end devices (functional approach). Due to the 
rapidly changing test conditions, especially concerning changing front 
end hardware, a flexible solution for the FIT was needed. This was 
realised by implementing a separate FIT for each frontend device like 
the iseg power supplies or the ELMB.
	
Additional FITs are being implemented to provide driver functionality 
for the WIENER power supplies and the regulator stations which still 
had to be monitored and controlled separately at the system test in 2005. 
Using the DDC data transfer (see below), a FIT for the BOC is currently 
under development.

\subsection{The System Integration Tool}
\label{SIT}	
The mapping of the channels to the detector devices (Figure 
\ref{fig:FIT-SIT}) is done by the SIT\footnote{SIT: \textbf{S}ystem 
\textbf{I}ntegration \textbf{T}ool}, which has a geographical structure. 
While relative small test setups only require a limited number of hardware 
connections (cabling), it will be impossible to manage all 35000 connections 
of the final detector using a functional approach. Once the physically 
connected hardware is connected to the DCS using the FIT, the SIT will 
create a virtual image of the detector inside the DCS. This image will 
then be used to navigate through the detector's geographical structure 
and to monitor and control the relevant data. The system therefore allows 
for operation of the DCS without deeper knowledge of the physical cabling, 
which was also tested at the system test.
	
\subsection{The DAQ-DCS Communication}
\label{DDC}
During the experiment, the DAQ\footnote{DAQ: \textbf{D}ata \textbf{A}c
\textbf{q}uisition} system will not only be responsible for taking physics 
data but also for starting and stopping of runs. On the other hand, DCS has 
to react correspondingly to ensure correct detector operation. Therefore it 
is necessary to synchronize the DAQ system with DCS.
	
For the synchronisation DDC\footnote{DDC: \textbf{D}AQ  \textbf{D}CS 
\textbf{C}ommunication} will be used\cite{WupDDC}. The DCS structure 
will follow the DAQ hierarchy, and on the bottom of the hierarchy the 
connection between DAQ and DCS will be established. The existing DDC 
package allows for the bidirectional transfer of data, for the transfer 
of messages from DCS to DAQ, and for the transfer of commands from DAQ to DCS.

\section{The System Test}
\label{SystemTest}
The aim of the established system test is to validate the concept of the 
overall design consisting of the power supply and detector control system 
as well as of the data acquisition system. Interactions between the various 
components were investigated. Of special interest are studies concerning 
noise and crosstalk, which could be introduced by the power supply system, 
the long services, or by the common mechanical structures. It consists of a 
bi-stave mounted inside a cooling box to ensure controlled environment 
(regarding temperature and humidity) and to protect the detector modules 
against light.
	
\subsection{Environmental Control}
\label{ST_PID}
To avoid condensation on the staves, the cooling box is flushed with 
nitrogen. Temperature and relative humidity in the cooling box are 
monitored and used to determine the dewpoint which is calculated with 
the Magnus \cite{Magnus} formula. The actual dewpoint is the input of 
a PVSS-script based on a PID controller\footnote{PID controller: 
\textbf{P}roportional-\textbf{I}ntegral-\textbf{D}erivative controller, 
standard feedback loop component in industrial control applications}. 
It determines the control variable dependent on the deviation from the 
set point, the sum of the deviations and the rate of change of the deviation. 
The control variable is the input current of a mass flow controller which 
regulates the nitrogen flow through the cooling box. The current of up to 
200 mA is provided by a modified SC-OLink card operating as a current source 
supply.
	
The aim was a fast falling of the dewpoint below the setpoint. Once it is 
below the dewpoint it should stay there without the need of constant, 
manual adjustment of the nitrogen flow. Further the flow rate should be 
minimized to reduce the hereby introduced heat.
	
With the PID control and additional implemented alert handling it is now 
possible to operate the system test without the need of an operator. 

\subsection{The Power Supply System}
\label{ST_PSS}
The supply of the system test setup is based on the components introduced 
in chapter \ref{PSS}. All power supplies and the regulator station are used. 
	Additionally one half stave is equipped with services as they will 
be used in the final experiment. Their characteristics, including their 
length, meet the properties as requested for ATLAS.

\begin{figure}[h]
  \includegraphics[angle=0, width=\textwidth]{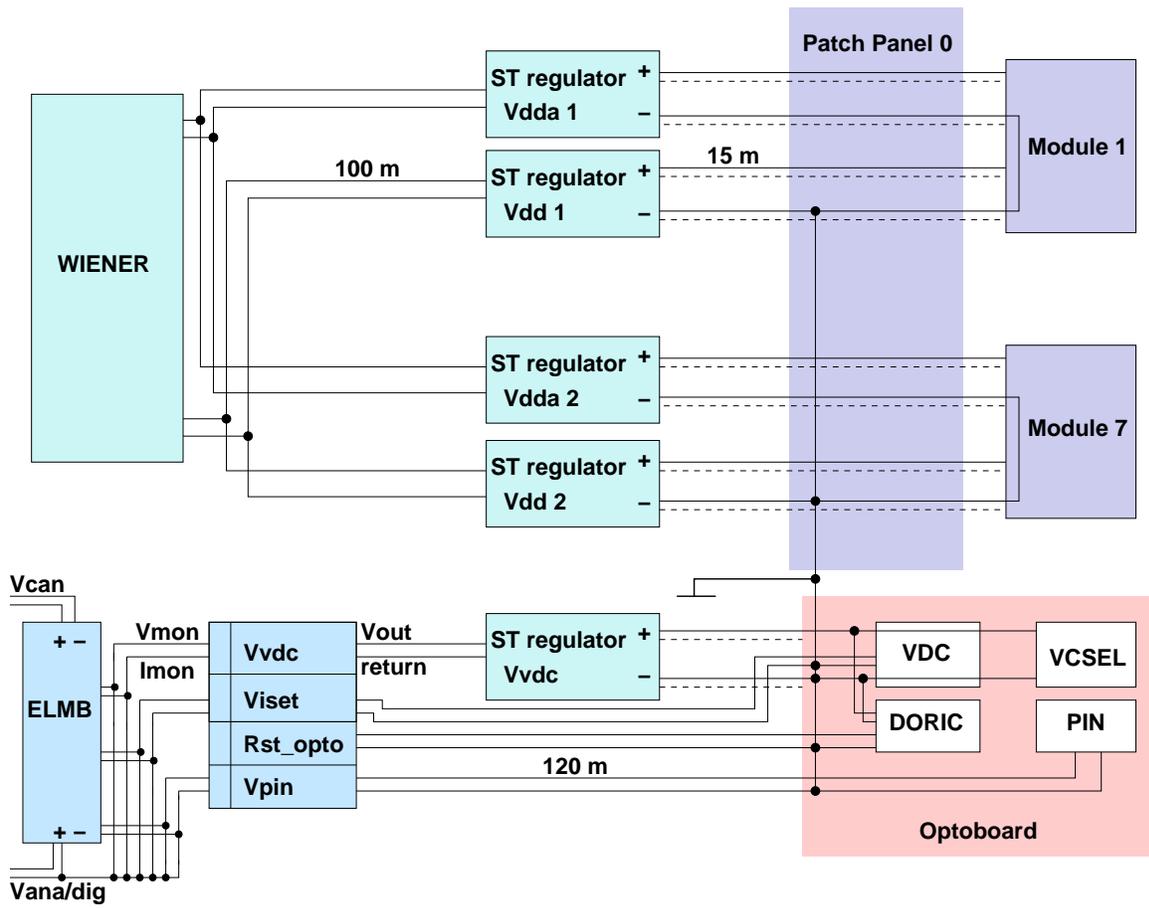}
  \caption{Grounding scheme}
  \label{fig:groundingScheme}
\end{figure}

The main difficulties during the installation and test of the system were 
caused by ground loops. It turned out that due to the long services 
potential differences could be built up which made the involved electronics 
non-operating (Figure \ref{fig:groundingScheme}). This effect made impact on 
the design of the regulator station and the SC-OLink. The monitoring circuits 
were therefor equipped with linear opto couplers which guarantee in all 
conditions a floating range larger than the required $\pm$\,10\,V.

\subsection{The Read out System}
\label{ST_DAQ}
The data path is also geared to the one used in the ATLAS experiment. An 
optical data transfer to the ead out crate is established. The DAQ software 
(STcontrol\footnote{STcontrol: \textbf{S}ystem \textbf{T}est 
\textbf{control}}) developed by the Bonn group is based on libraries of 
the ATLAS DAQ, as they are available at the moment.
	
\subsection{Procedure of the System Test}
\label{Procedure}
To evaluate the performance of the whole setup, several tests for the 
digital and the analog part of the the frontend electronics were enforced 
for each detector module in several states. Starting at the assembly of a 
detector module down to the final state when mounted on the stave and 
integrated in the system test environment, all these tests together allow 
a rating of the detector module's performance.

Altogether we got results from four different tests for comparison. The 
first two are using the well understood laboratory setup, as used during 
production qualification. It is based on laboratory power supplies with 
remote sensing, short services and an electrical read out. 

\begin{description}
\item [Module Assembly]- the individual detector module is measured directly 
after its assembly with the laboratory setup
\item [Stave Assembly]- after mounting all detector modules on the stave the 
perfomance of each detector module is verified using again the laboratory setup
\end{description}

The last two results, obtained in the system test (ST), are all based on the 
system foreseen for the final experiment, which was built up for the first 
time in its complexity. The power supply system and the optical ead out chain 
as previously described were used together with long, realistic services.
\begin{description}
\item [ST separate operation]- each detector module of the stave is operated 
individually using the final setup
\item [ST parallel operation]- parallel operating of six detector modules 
on a stave using the final setup
\end{description}

\subsection{Results}
\label{Results}

All data scans were performed with the configuration of the read out chips 
determined during the assembly tests. This avoids impacts on the results 
due to differences in the quality of the tuning. The disadvantage of this 
procedure is the temperature dependency of the results of the analog scans. 
	
\subsubsection{Digital}
The digital part showed no effect to the different states, but it was never 
considered as the critical part sensitive to crosstalk.
	
\subsubsection{Analog}
\begin{figure}[htb]
  \includegraphics[angle=-90,width=\columnwidth]{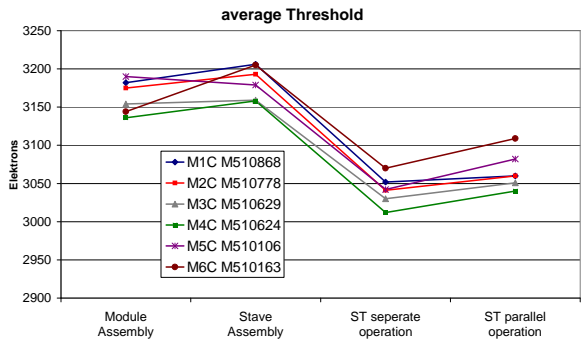}
  \caption{Comparison of the threshold for the different states}
  \label{fig:Threshold}
\end{figure}

The behaviour of the threshold (Figure~\ref{fig:Threshold}) shows a 
correlated behaviour for all detector modules for each state. Obviously 
the threshold is also correlated to the operating temperatures (Module 
Assembly @ 25\,$^\circ$C, Stave Assembly @ 27\,$^\circ$C, ST separate 
operation @ 18\,$^\circ$C and ST parallel operation @ 20\,$^\circ$C), 
it raises with the temperature.

\begin{figure}[htb]
  \includegraphics[angle=-90,width=\columnwidth]{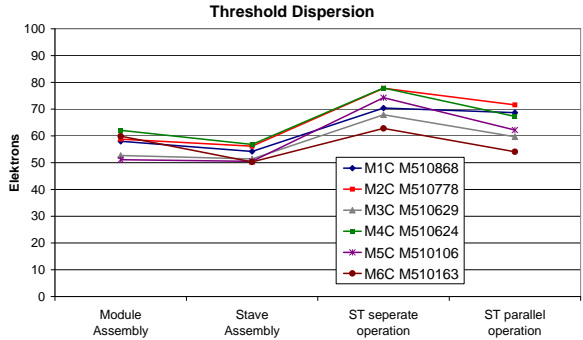}
  \caption{Comparison of the threshold dispersion for the different states}
  \label{fig:ThrDisp}
\end{figure}

As only one configuration was used and the threshold of the pixel cells are 
non-uniformly temperature dependent, the threshold dispersion 
(Figure~\ref{fig:ThrDisp}) increases with the difference between the 
tuning temperature and the operation temperature.

\begin{figure}[htb]
  \includegraphics[angle=-90,width=\columnwidth]{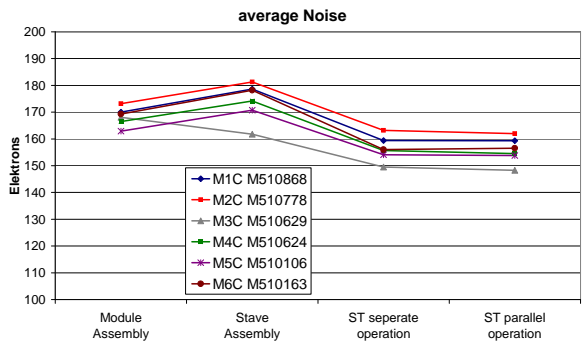}
  \caption{Comparison of the noise for the different states}
  \label{fig:Noise}
\end{figure}

The noise (Figure \ref{fig:Noise}) is only dependent on the absolute 
temperature of the electronics and shows no other effect as would be 
expected for a potential crosstalk. 

\section{Summary and Outlook}
\label{Summary}
A system test utilizing powering and services as foreseen to be used for the
ATLAS experiment has been constructed, and has provided valuable first 
operational experience.  The detector control system successfully ensured
stable operation of the system.  The grounding scheme envisioned for the 
experiment has been studied extensively, resulting in a modified design for
some components of the power supply system.

The experience gained thus far from operating the system test setup indicates
that the system of power supplies foreseen for the experiment, as well as
the optical readout chain, operate as desired and introduce no problematic
effects to the system.  This has been verified by operating single modules
as well as several modules in parallel, where no deleterious influence has 
been observed.

The only observable influence was due to the different temperatures of 
operation.  To compensate for the correlation of the threshold dispersion, 
the detector modules will be tuned again for a set of different temperatures. 
This should allow a direct comparison for one configuration.
	
Another remaining objective is to scale up the number of detector modules 
operated in parallel to check for crosstalk due to the additional hardware 
and services introduced, as well as to demonstrate the scalability of the 
readout system and the detector control system.

\section{Acknowledgments}
\label{Acknowledgements}
For building up and debugging of the system test setup, the knowledge and 
encouragement of all participating institutes - especially of the ATLAS 
pixel community - were necessary. The cooperation was highly productive. 
Thanks to all involved colleagues.



\onecolumn

\end{document}